\documentclass[reprint]{revtex4-2}
\usepackage{graphicx}
\usepackage{epsfig}
\usepackage{epstopdf}
\usepackage{amsfonts}
\usepackage{amssymb}
\usepackage{amsbsy}
\usepackage{amsmath}
\usepackage{mathrsfs}
\usepackage{latexsym}
\usepackage{natbib}
\usepackage{bm}
\usepackage{color}
\usepackage{braket}
\usepackage{slashed}
\usepackage{pgfplots}

\DeclareMathOperator{\arcsinh}{asinh}

\usepackage{tikz}
\usetikzlibrary{shapes,arrows,shadows}

\def\x{\mathbf{x}}

\def\k{\mathrm{k}}

\def\mr{\mathcal{M}}

\def\muD{{\bf\mu_D}}

\begin{document}

\author{Golam Mortuza Hossain}
\email{ghossain@iiserkol.ac.in}

\affiliation{ Department of Physical Sciences, 
Indian Institute of Science Education and Research Kolkata,
Mohanpur - 741 246, WB, India }

\author{Susobhan Mandal}
\email{sm12ms085@gmail.com}

\affiliation{Department of Physics, 
Indian Institute of Technology Bombay,
Mumbai - 400076, India }


\date{\today}

\title{Effects of magnetic field on the equation of state in 
curved spacetime of a neutron star}

\begin{abstract}

Neutron stars are known to have strong magnetic fields reaching as high 
as $10^{15}$ Gauss, besides having strongly curved interior spacetime. So for 
computing an equation of state for neutron-star matter, the effect of magnetic 
field as well as curved spacetime should be taken into account. In this article, 
we compute the equation of state for an ensemble of degenerate fermions in the 
curved spacetime of a neutron star in presence of a magnetic field. We show 
that the effect of curved spacetime on the equation of state is relatively 
stronger than the effect of observed strengths of magnetic field. Besides, a 
thin layer containing only spin-up neutrons is shown to form at the boundary of 
a degenerate neutron star.
\end{abstract}

\maketitle

\section{Introduction}

The astrophysical data suggest that the surface magnetic field of a typical 
neutron star is around $10^{11} - 10^{13}$ Gauss, whereas the internal field 
strength can reach up to $10^{15}$ Gauss or even higher 
\cite{bignami2003magnetic, igoshev2021evolution, enoto2019observational}. The 
dominant matter constituent of neutron stars are believed to be charge-less 
neutrons. However, they interact with a magnetic field through the non-minimal 
Pauli-Dirac gauge coupling due to their intrinsic magnetic moment. Therefore, 
the presence of a strong magnetic field is expected to play an important role in 
determining thermodynamic properties of matter present inside the neutron 
stars. At the same time, neutron stars are also expected to contain 
electrically charged particles like protons and electrons. These charged 
particles directly interact with a magnetic field and form the well-known 
Landau levels quantum mechanically. These Landau levels are bound states of 
charged particles. Therefore, it is important to study their roles in 
computation of the fermionic degeneracy pressure which makes compact stars 
such as neutron stars stable against the gravitational collapse. The 
thermodynamic properties of a gas of charged particles under an external 
magnetic field in the Minkowski spacetime, have been studied earlier in 
\cite{canuto1968quantum, canuto1968thermodynamic, canuto1968magnetic}.

However, recent articles \cite{hossain2021equation, hossain2021higher} show that 
the curved background geometry of a neutron star also plays a crucial role in 
determining the properties of the equation of state (EOS) of the matter present 
inside the star. In particular, metric-dependent gravitational time-dilation 
effect leads to an enhancement of the stiffness of the EOS of matter. 
Consequently, such an EOS, referred to as the curved EOS, leads to an 
enhancement of the mass limits of neutron stars \cite{hossain2021higher}. We 
have mentioned that the observed neutron stars are known to have strong magnetic 
fields. Therefore, it is important to take the magnetic field into account while 
computing the EOS for a neutron star in its curved spacetime.

The key idea that we use for computing the EOS here is the lesson from 
Einstein's general relativity that even in a curved spacetime one can always 
find a set of local coordinates in which the spacetime metric appears to be 
\emph{locally} flat. However, it is unlike the usage of a \emph{globally} flat 
spacetime which is commonly used in the literature to compute the EOS, referred 
to as the flat EOS, for neutron stars. Subsequently, we employ the methods of 
thermal quantum field theory to compute the EOS, as pioneered by Matsubara 
\cite{matsubara1955new}. The result derived here shows that for an ensemble of 
charge-less neutrons the magnetic field and the gravitational time-dilation both 
leads the EOS to become stiffer whereas for an ensemble of charged fermions the 
magnetic field makes the EOS softer due to formation of the Landau levels. 
However, the changes of stiffness of the EOS due to the gravitational time 
dilation effect is relatively stronger than the changes due to the observed 
strengths of magnetic field.

\section{Interior spacetime}

In the presence of an axial magnetic field, the interior spacetime of a neutron 
star can be modelled by an axially symmetric spacetime. On the other hand, the 
spacetime metric of a slowly rotating star that preserve axial symmetry can be 
represented, in the \emph{natural units} $c = \hbar =1$, by the following 
invariant line element \cite{hartle1967slowly, hartle1968slowly}
\begin{equation}\label{SRMetric}
ds^2 = - e^{2\Phi}dt^2 + e^{2\nu} dr^2 + 
r^2[d\theta^2 +  \sin^2\theta (d\varphi-\omega dt)^2] ~,
\end{equation}
where $\omega = \omega(r)$ is the acquired angular velocity of a freely-falling 
observer from infinity, a phenomena referred to as the \emph{dragging} of 
inertial frames. On the other hand, the radial variation of the metric function 
$\Phi = \Phi(r)$ leads to the phenomena of gravitational \emph{time dilation}. 
We note that in absence of the frame-dragging angular velocity $\omega$, the 
spacetime metric (\ref{SRMetric}) represents a spherically symmetric spacetime.

The contribution of inertial frame-dragging on the EOS is controlled by a 
dimensionless ratio $(\omega/m)$ and if we consider $m$ to be the mass of 
neutrons then even for a rapidly rotating millisecond pulsar the dimensionless 
ratio is vanishingly small as $(\omega/m) \sim 10^{-22}$ 
\cite{hossain2022equation}. Nevertheless, similar to the magnetic field $B$, the 
frame-dragging angular velocity $\omega$ couples to the spin-component of the 
Dirac field. However, as we have argued that the effect of inertial 
frame-dragging on the EOS is extremely small. So for computation of the EOS in 
the presence of a magnetic field, we neglect the inertial frame-dragging and we 
take into account only the effect of gravitational time-dilation, by considering 
the following invariant line element 
\begin{equation}\label{SphericalMetric}
ds^2 = - e^{2\Phi}dt^2 + e^{2\nu} dr^2 + r^2[d\theta^2 
+ \sin^2\theta d\varphi^2] ~,
\end{equation}
which essentially represents a spherically symmetric spacetime.

\section{Anisotropic Pressure due to Magnetic field}

In order to study the \emph{interior} spacetime, here we consider the stellar 
matter to be described by a perfect fluid with the stress-energy tensor
\begin{equation}\label{StressEnergyTensorPerfectFluid}
T^{M}_{\mu\nu} = (\rho + P) u_{\mu} u_{\nu} + P g_{\mu\nu} ~,
\end{equation} 
where $u^{\mu}$ is the 4-velocity of the stellar fluid satisfying 
$u_{\mu}u^{\mu}= -1$, $\rho$ is the energy density and $P$ is the pressure of 
the fluid. On the other hand, the stress-energy tensor associated with the 
electromagnetic field is given by
\begin{equation}
T^{E}_{\mu\nu} = \frac{1}{\mu_0} \left[ F_{\mu\alpha}F_{ \ \nu}^{\alpha} 
- \frac{1}{4}g_{\mu\nu}F_{\alpha\beta} F^{\alpha\beta}  \right] ~,
\end{equation}
where $\mu_0$ is the magnetic permeability of vacuum and $F_{\mu\nu} = 
\partial_{\mu}\mathcal{A}_{\nu} - \partial_{\nu} \mathcal{A}_{\mu}$ is the
electromagnetic field tensor whose indices are contracted with respect to the 
spacetime metric. Therefore, the total stress-energy tensor $T_{\mu\nu} = 
T^{M}_{\mu\nu} + T^{E}_{\mu\nu}$, can be expressed in the following form
\begin{equation}
T_{ \ \nu}^{\mu} = \text{diag}( - \rho, P_{r}, P_{t}, P_{t}) ~.
\end{equation}
We note that due to the presence of a magnetic field the total radial 
pressure $P_{r}$ and total tangential pressure $P_{t}$ differ from each other.
On the other hand total energy density $\rho$ now includes the contributions 
from both stellar fluid as well as from the magnetic field. The $tt$ and $rr$ 
components of Einstein's equation $G_{ \ \nu}^{\mu} = 8\pi G T_{ \ \nu}^{\mu}$ 
corresponding to the metric (\ref{SphericalMetric}), lead to 
the following equations
\begin{equation}
\begin{split}
8\pi G r^{2}\rho & = e^{-2\nu}(2r\nu' - 1) + 1 ~,\\
8\pi G r^{2}P_{r} & = e^{-2\nu}(2r\Phi' + 1) - 1 ~.
\end{split}
\end{equation}
By considering $e^{2\nu} = 1/(1 - 2G\mathcal{M}/r)$, we obtain the equations
\begin{equation}\label{TOVEqn}
\frac{d\Phi}{dr} = \frac{G(\mr + 4\pi r^3 P_r)}{r(r - 2 G \mr)} ~~,~~
\frac{d\mathcal{M}}{dr} = 4\pi r^{2}\rho ~.
\end{equation}   
The conservation of stress-energy tensor leads to the equation as follows
\begin{equation}
\frac{dP_{r}}{dr} = \frac{2}{r}(P_{t} - P_{r}) - (P_{r} 
+ \rho)\frac{d\Phi}{dr} ~.
\end{equation}
Additionally, we also have a second-order differential equation which follows 
from $G_{ \ \theta}^{\theta} = 8\pi G T_{ \ \theta}^{\theta}$ equation and is 
given by
\begin{equation}\label{GThetaThetaEq}
8\pi G \ P_{t} = e^{-2\nu}\Big[\frac{d^{2}\Phi}{dr^{2}} - \frac{d\Phi}{dr}
\frac{d\nu}{dr} + \left(\frac{d\Phi}{dr}\right)^{2} + \frac{1}{r}\left(\frac{d\Phi}{dr} 
- \frac{d\nu}{dr}\right)\Big] ~.
\end{equation}
However, we note that the equation (\ref{GThetaThetaEq}) is not an independent 
equation and it can be obtained from the conservation equation and 
$rr$-component of Einstein's equation.

For a detailed study on the anisotropic spherical star in general relativity, 
see \cite{bowers1974anisotropic}. Nevertheless, we note that for an 
axial magnetic field $B$, the pressure components $P_t$ and $P_r$ would differ 
from each other by the terms of $\mathcal{O}(B^2)$. So for the cases where 
$\mathcal{O}(B^2)$ terms are negligible, the total pressure can be considered to 
be isotropic. We shall see later that the observed field strength of even 
$10^{15}$ Gauss is significantly smaller compared to the characteristic field 
strength $B_c \approx 10^{20}$ Gauss associated with the nucleons. It allows us 
to neglect the terms of $\mathcal{O}(B^2)$ which in turn permits the exterior 
metric to be described by the Schwarzschild metric such that metric function 
$\Phi$ is subject to the boundary condition $e^{2\Phi(R)} = 1 - 2GM/R$. For a 
typical neutron star having mass $M = 1 M_{\odot}$ and radius $R=10$ km, the 
metric function $\Phi(R) \simeq -0.17$. Further, it follows from the equation 
(\ref{TOVEqn}) that the values of $\Phi$ inside the star are lower than 
$\Phi(R)$ as $(d\Phi/dr)$ is positive definite.

\section{Local thermal equilibrium}

Due to the hydrostatic equilibrium, the thermodynamic properties such as the 
pressure, the energy density vary radially inside a star. On the other hand, 
these thermodynamic properties are required to be uniform within a given 
thermodynamical system in equilibrium. Nevertheless, these two seemingly 
disparate aspects can be reconciled by introducing the concept of \emph{local} 
thermodynamical equilibrium inside the star.

In order to ensure the conditions for  local thermodynamic equilibrium 
inside a star, we can choose a sufficiently small region but containing large 
number of degrees of freedom. Inside this small region the metric variations 
can be neglected. For definiteness, we chose a box-shaped small region whose 
center is located at say $r=r_{0}$. By following the coordinate transformations 
given in \cite{hossain2021equation}, namely
$x = e^{\nu(r_0)} r \sin \bar{\theta} \cos\phi$, 
$y = e^{\nu(r_0)} r \sin \bar{\theta} \sin\phi$, and
$z = e^{\nu(r_0)} r \cos \bar{\theta}$ along with $\bar{\theta} =  
e^{-\nu(r_0)}\theta$ for small $\theta$, we can reduce the metric 
(\ref{SphericalMetric}) to the following form 
\begin{equation}\label{MetricInBox}
ds^{2} = -e^{2\Phi(r_0)}dt^{2} + dx^{2} + dy^{2} + dz^{2},
\end{equation}
in a locally Cartesian coordinates. The metric within the box 
(\ref{MetricInBox}) contains the information about the metric function $\Phi = 
\Phi(r_0)$, in contrast to the usage of a globally flat spacetime for computing 
the matter EOS in the literature \cite{shen2002complete, douchin2001unified, 
lattimer2016equation, tolos2016equation, ozel2016dense, katayama2012equation}. 
The metric function $\Phi$ is treated here as a constant within the \emph{scale 
of the box}, which is sufficient to describe the microscopic physics of the
constituent particles. However, the metric function $\Phi$ varies at the 
\emph{scale of the star}, as governed by the equations (\ref{TOVEqn}).

\section{Neutrons in an external magnetic field}

Neutrons are electrically neutral particles, hence, they do not couple
\emph{minimally} to the gauge field associated with the external magnetic 
field. However, neutrons possess a magnetic dipole moment due to their internal 
quark degrees of freedom. Consequently, under an external magnetic field, 
neutrons couple to the gauge field \emph{non-minimally} through the Pauli-Dirac 
interaction. The corresponding action is given by
\begin{equation}\label{action of neutrons}
S = -\int\sqrt{-g}d^{4}x \bar{\psi} \Big[ie_{ \ a}^{\mu}
\gamma^{a}\mathcal{D}_{\mu} + m - \frac{\muD}{2}\sigma^{\mu\nu}
F_{\mu\nu}\Big]\psi ~,
\end{equation} 
where spinor field $\psi$ represents the neutrons with mass $m$ 
and $\bar{\psi} = \psi^{\dagger}\gamma^0$ being its Dirac adjoint. The 
\emph{tetrad} components ${e^{\mu}}_a$ are defined as $g_{\mu\nu} {e^{\mu}}_a 
{e^{\nu}}_b = \eta_{ab}$ where  $g_{\mu\nu}$ is the spacetime metric whereas  
$\eta_{ab} = diag(-1,1,1,1)$ is the Minkowski metric. The spin-covariant 
derivative is defined as $\mathcal{D}_{\mu}\psi \equiv \partial_{\mu}\psi + 
\Gamma_{\mu}\psi $ where spin connection $\Gamma_{\mu}$ is given by
\begin{equation}\label{OmegaMuabDef}
\Gamma_{\mu}  =  -\tfrac{1}{8} \eta_{ac} {e_{\nu}}^c 
(\partial_{\mu} {e^{\nu}}_b + \Gamma^{\nu}_{\mu\sigma} {e^{\sigma}}_b )
~[\gamma^{a}, \gamma^{b}]  ~,
\end{equation}
with $\Gamma^{\nu}_{\mu\beta}$ being the Christoffel connections. The Dirac 
matrices $\gamma^{a}$ satisfy the Clifford algebra $\{\gamma^{a},\gamma^{b}\} = 
- 2\eta^{ab} \mathbb{I}$ together with the relations $(\gamma^0)^2 = 
\mathbb{I}$ and $(\gamma^k)^2 = -\mathbb{I}$ for $k=1,2,3$. In the Pauli-Dirac 
interaction term, $\muD$ denotes the \emph{magnitude} of magnetic moment of 
neutrons and $\sigma^{\mu\nu} =  \frac{i}{2} {e^{\mu}}_a {e^{\nu}}_b [\gamma^a, 
\gamma^b]$.

\subsection{Partition function}

In order to compute the partition function around a given a point inside the 
star, we consider a small region around it where the metric can be reduced to 
the form (\ref{MetricInBox}). Within this box-shaped region the tetrad 
components can be expressed as $e_{\ 0}^{t} = e^{-\Phi}, \ e_{ \ 1}^{x} = 
e_{\ 2}^{y} = e_{ \ 3}^{z} = 1$. Consequently, the spin-connection within the 
box vanishes \emph{i.e.} $\Gamma_{\mu} = 0$. Additionally, we choose the 
magnetic field to be in the $z$-direction with its field strength being $B$. 
For such a magnetic field the gauge field components can be chosen as 
$\mathcal{A}_{\mu} = (0,0,B x,0)$. Therefore, within the Box, the action 
(\ref{action of neutrons}) reduces to the following form
\begin{equation}
S = -\int d^{4}x \bar{\psi} \Big[i\gamma^{0}\partial_{t} + e^{\Phi}
(i\gamma^{k}\partial_{k} + m) - e^{\Phi}\muD B\Sigma_3\Big]\psi ~,
\end{equation}
where $\Sigma_3 = \tfrac{i}{2}[\gamma^{1},\gamma^{2}] = \sigma_3 \otimes 
\mathbb{I}_2$ and $\sigma_3$ is the Pauli matrix associated with the spin 
operator along $z$-direction.

In functional integral approach, the partition function can be expressed as 
$\mathcal{Z} = \int\mathcal{D} \bar{\psi} \mathcal{D}\psi \ e^{-S^{\beta}}$ 
by using the coherent states of the Grassmann fields \cite{laine2016basics, 
kapusta1989finite, das1997finite}. Here $S^{\beta}$ denotes the Euclidean action 
obtained obtained through the Wick rotation $t\rightarrow -i\tau$. By following 
the approach as given in \cite{hossain2021equation,hossain2022equation}, we can 
express the Euclidean action as
\begin{eqnarray}
S^{\beta} = \int_{0}^{\beta}d\tau\int d^{3}x \bar{\psi}\big[
&-& \gamma^{0}(\partial_{\tau} + \mu + e^{\Phi}\muD B \gamma^0 \Sigma_3) 
\nonumber \\ \label{EuAction0}
&+& e^{\Phi} (i\gamma^{k}\partial_{k} + m)\big]\psi .
\end{eqnarray}
The equilibrium temperature $T$ of the system leads to the following 
\emph{anti-periodic} boundary condition on the Dirac field 
\begin{equation}\label{FermionicBoundaryCondition}
\psi(\tau,\x) = -\psi(\tau+\beta,\x)  ~,
\end{equation}
where $\beta = 1/(k_B T)$ with $k_B$ being the Boltzmann constant. By using the 
Matsubara frequencies $\omega_l = (2l+1) \pi/\beta$ where $l$ is an integer, we 
can express the field $\psi$ in the Fourier domain as
\begin{equation}\label{FermionicFourier}
\psi(\tau,\x) = \frac{1}{\sqrt{V}} \sum_{l,\k} ~e^{-i(\omega_l\tau + 
\k\cdot\x)} \tilde{\psi}(l,\k)  ~,
\end{equation}
where volume of the box is now $V = \int d^3x \sqrt{-\eta}$. The equation 
(\ref{FermionicFourier}) then leads the action (\ref{EuAction0}) to become
\begin{equation}\label{EuAction0IIN}
S^{\beta} = \sum_{l,\k} ~\bar{\tilde{\psi}}~\beta
\left[ \slashed{p} + \bar{m} \right]  \tilde{\psi} ~,
\end{equation}
where $\slashed{p} = \gamma^{0}(i\omega_l - \mu -  e^{\Phi}\muD B \gamma^0 
\Sigma_3) + \gamma^{k} (\k_k e^{\Phi})$ and $\bar{m} =  m e^{\Phi}$.
Using the results of Gaussian integral over the Grassmann fields and the Dirac 
representation of $\gamma^a$ matrices, one can evaluate the partition function 
$\mathcal{Z}$ for the \emph{particle} sector as
\begin{equation}\label{PartionFunctionMus0}
\ln\mathcal{Z} = \sum_{s = \pm}\sum_{\k}\ln\left(1 + e^{\beta(\mu_{s} - 
\varepsilon)}\right) ~,
\end{equation}
where $\varepsilon^2 = \varepsilon(\k)^2 = e^{\Phi}(\k^2 + m^2)$
and the modified chemical potential associated with the different spins of 
neutrons are
\begin{equation}\label{MuSDefinition}
\mu_{s} = \mu + s e^{\Phi}\muD B ~.
\end{equation}
In general, the presence of a magnetic field makes the dispersion relation 
anisotropic in the momentum space. As a result, the Fermi-surface is no longer 
spherical in nature, rather it becomes an ellipsoid. However, for $(\muD 
B/m)\ll 1$ limit (which is typically the case inside a neutron star) and 
neglecting the anisotropy due to the fact $(k_z/m) \ll 1$, we get the following 
two dispersions
\begin{equation}
\omega = e^{\Phi} (\varepsilon \pm \muD B) ~,
\end{equation}
which are two shifted spheres. In the thermodynamic limit, the summation over 
$\k$ in the equation (\ref{PartionFunctionMus0}) can be expressed as an integral 
over the momentum space that results in the following expression of the 
partition function
\begin{equation}\label{NeutronsLogZfinal}
\ln\mathcal{Z} = \sum_{s = \pm}\frac{e^{-3\Phi}\beta V}{48\pi^{2}}
\Big[2\mu_{s}\mu_{sm}^{3} - 3\bar{m}^{2}\bar{\mu}_{sm}^{2}\Big] ~,
\end{equation}
where $\mu_{sm} = \sqrt{\mu_{s}^{2} - \bar{m}^{2}}$ and $\bar{\mu}_{sm}^{2} = 
\mu_{s}\mu_{sm} - \bar{m}^{2}\arcsinh(\mu_{sm}/\bar{m})$. In arriving at 
the expression (\ref{NeutronsLogZfinal}), we have neglected the temperature 
corrections of $\mathcal{O}((\beta\mu)^{-2})$, given a degenerate star is 
characterized by the condition $(\beta\mu)\gg 1$. Additionally, we have 
omitted formally divergent zero-point energy terms.

\subsection{Pressure and energy density}

We can compute number density of neutrons from the partition function 
(\ref{NeutronsLogZfinal}) as $n = (\beta V)^{-1} (\partial 
\ln\mathcal{Z})/(\partial\mu)$ which leads to 
\begin{equation}\label{NumberDensityPM}
n = n_{+} + n_{-} ~,~~  \text{with} ~~
n_{\pm} = \frac{e^{-3\Phi}}{6\pi^{2}} \mu_{\pm m}^{3}  ~.
\end{equation}
The equation (\ref{NumberDensityPM}) can be used to express the modified 
chemical potentials in terms of the number densities of spin-up and spin-down 
neutrons respectively as $\mu_{\pm} = m e^{\Phi} \sqrt{(bn_{\pm})^{2/3} + 1}$ 
where $b = (6\pi^2)/m^3)$. We should mention here that $\mu_{\pm}$ can be 
equivalently treated as independent variables in places of $\mu$ and $B$. The 
equation (\ref{MuSDefinition}) then leads to the following relation
\begin{equation}\label{NplusNMinusRelation}
\sqrt{(bn_{+})^{2/3} + 1} - \sqrt{(bn_{-})^{2/3} + 1} = 
2 \left(\frac{B}{B_c}\right) ~,
\end{equation}
where the constant $B_c = (m/\muD) \approx 10^{20}$ Gauss. The constant 
$B_c$ here signifies the characteristic scale of magnetic field associated with 
neutrons.

For a grand canonical ensemble, we can compute total pressure from the 
partition function (\ref{NeutronsLogZfinal}) as $P = (\beta V)^{-1} 
\ln\mathcal{Z} = P_{+} + P_{-}$ where the pressure components associated 
with the different spins of neutrons are
\begin{eqnarray}\label{PressurePM}
P_{\pm} = e^{\Phi} \frac{m^4}{48\pi^2} && \left[ \sqrt{(b n_{\pm})^{2/3} + 1}
\left\{2(b n_{\pm}) - 3(b n_{\pm})^{1/3} \right\} \nonumber \right. \\
&& + \left. ~3 \arcsinh\left\{ (b n_{\pm})^{1/3} \right\} \right]  ~.
\end{eqnarray}
For the metric (\ref{MetricInBox}), the 4-velocity vector in the box 
corresponding to the perfect fluid form of the stellar fluid 
(\ref{StressEnergyTensorPerfectFluid}) can be expressed as $u^{\mu} = 
e^{-\Phi}(1, 0, 0, 0)$ along with  its co-vector $u_{\mu} = e^{\Phi}(-1, 0, 0, 
0)$. Consequently, the energy density $\rho$ can be expressed in terms of the 
partition function as $(\rho -\mu n)V = - 
(\partial\ln\mathcal{Z}/\partial\beta)$ \cite{hossain2022equation} leading to 
$\rho = \rho_{+} + \rho_{-}$ where
\begin{equation}\label{EnergyDensityPM}
\rho_{\pm} = - P_{\pm} + e^{\Phi}\frac{m^4}{6\pi^2} (b n_{\pm})
\sqrt{(b n_{\pm})^{2/3} + 1} ~. 
\end{equation}
In the limit $B\to 0$, we note that $\mu_{+}=\mu_{-}$. Consequently, in this 
limit total pressure $P$ and total energy density $\rho$ reduce to the 
expressions of pressure and energy density for an ensemble of non-interacting 
degenerate neutrons as expected. For a non-zero $B$, it can be checked that the 
corrections to the total pressure $P$ and total energy density $\rho$ are of 
$\mathcal{O}(B^2)$.

\begin{figure}
\includegraphics[height = 7cm, width = 9cm]{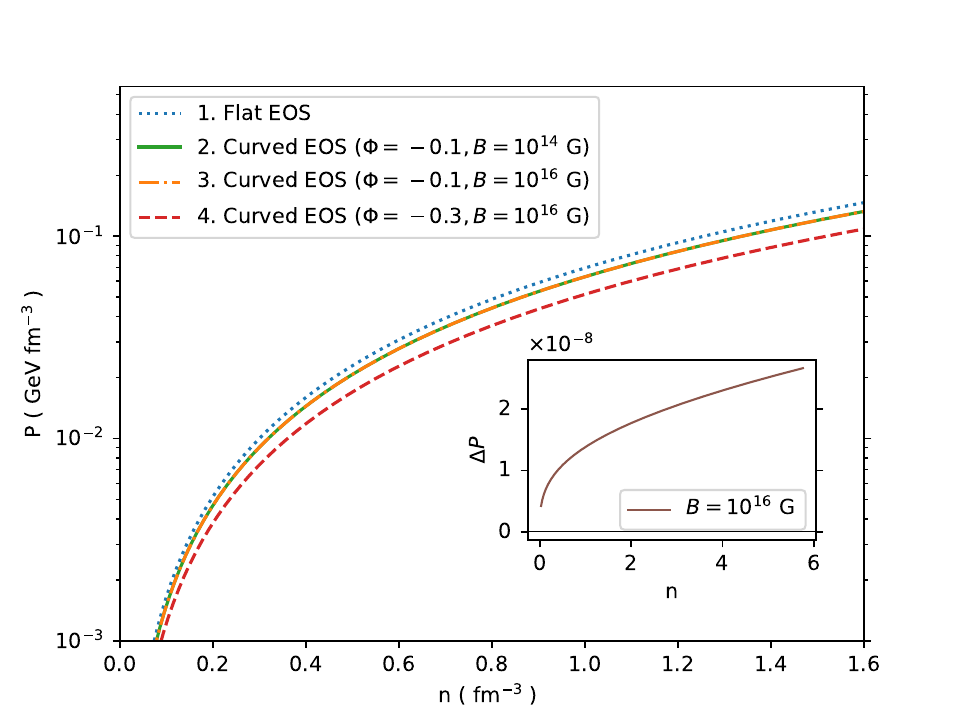}
\caption{Plot of the pressure $P$ exerted by an ensemble of non-interacting 
degenerate neutrons under an external magnetic field $B$ as a function of 
neutron number density $n$ for different kinematical values of metric function 
$\Phi$. The curves $2$ and $3$ with different plausible values of magnetic field 
$B$ are nearly indistinguishable. The change of pressure due to the magnetic 
field $B$ \emph{i.e.} $\Delta P \equiv P_{B} - P_{B=0}$ is shown in the inset 
plot using same units as the  main plot. It shows that effect of magnetic field 
on the pressure is rather small. In contrast, the effect of gravitational 
time-dilation on the pressure, described by $\Phi$, is considerably large.}
\label{fig:pressure_comparison_N}
\end{figure}

\begin{figure}
\includegraphics[height = 7cm, width = 9cm]{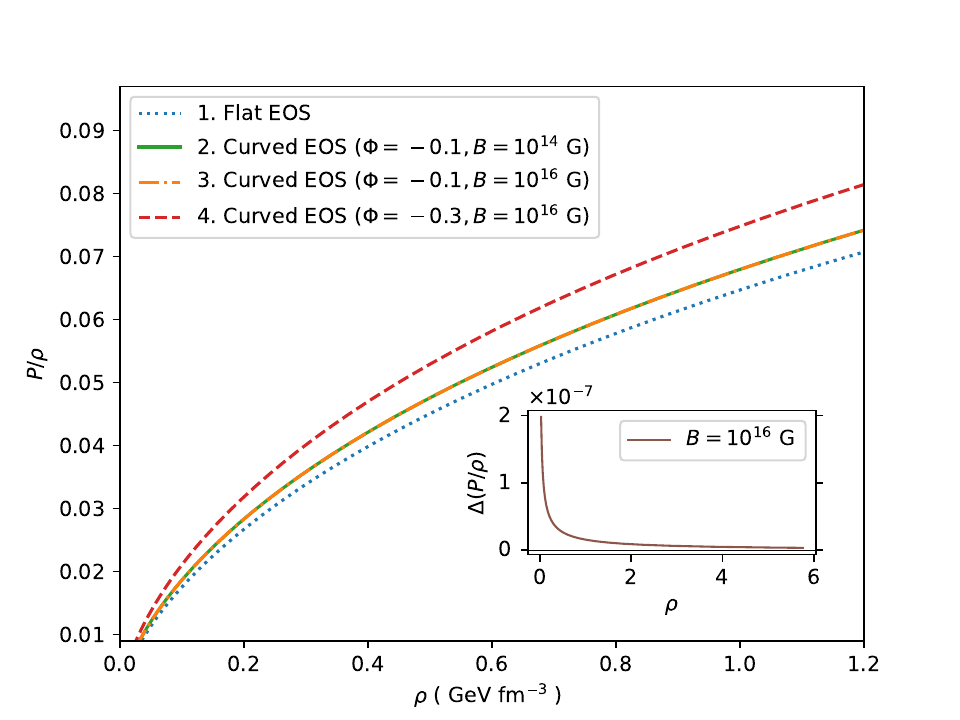}
\caption{Plot of the ratio $P/\rho$ as a function of energy density $\rho$ 
for different values of magnetic field $B$ and metric function $\Phi$. As 
earlier, the curves $2$ and $3$ with different values of magnetic field $B$ are 
nearly indistinguishable. The change of the ratio $P/\rho$ due to the magnetic 
field $B$ \emph{i.e.} $\Delta (P/\rho) \equiv (P/\rho)_{B} - (P/\rho)_{B=0}$ is 
shown in the inset plot using same unit for $\rho$ as the main plot. As earlier, 
we note that the effect of gravitational time-dilation on the EOS is relatively 
stronger compared to the effect of an external magnetic field. Nevertheless, the 
EOS here becomes stiffer due to the effects of both the gravitational 
time-dilation and an external magnetic field.}
\label{fig:eos_ratio_N}
\end{figure}

The behaviour of total pressure as a function of number density for 
different values of the magnetic field and the metric function is plotted in 
the FIG. \ref{fig:pressure_comparison_N}. On the other hand, the FIG. 
\ref{fig:eos_ratio_N} shows the behaviour of pressure, energy density ratio as 
a function of energy density and it shows the stiffening of the 
EOS due to the effects of both magnetic field and gravitational time dilation.

\subsection{Magnetic moment of a neutron star}

We note from the equation (\ref{NplusNMinusRelation}) that for a non-zero value 
of magnetic field $B$, there is a population difference between different spins 
of neutrons. As a result, number densities of spin-up and spin-down neutrons, 
$n_{+}$ and $n_{-}$ respectively, cannot vanish simultaneously at the boundary 
of the star, and namely, $n_{-}$ vanishes earlier inside the neutron star. In 
particular, when $n_{-}$ becomes zero then we obtain 
\begin{equation}
n_{+} = \frac{8}{b} \left[\frac{B}{B_c} \left(1 + \frac{B}{B_c} \right) 
\right]^{3/2} ~.
\end{equation}
Consequently, due to the presence of a magnetic field there exists a thin layer 
at the boundary of a degenerate neutron star which contains only spin-up 
neutrons. In turn, the neutron star as a whole would acquire a net magnetic 
moment which would then naturally lead to an accretion of charge particles 
surrounding the neutron star. In the case of rotating neutron stars, analogous 
thin layer containing only one kind of spins has been reported earlier 
\cite{SeedMagnetism2022} where it arises due to the dragging of inertial 
frames.

\section{Charged fermions in an external magnetic field}

We have mentioned earlier that primary constituents of a neutron star 
are believed to be neutrons. However, a neutron star is also expected to have
a smaller fraction of electrically charged fermions such as protons and 
electrons. Unlike neutrons,  charged fermions couple \emph{minimally} with 
the gauge field associated with an external magnetic field. We shall, however, 
ignore the contributions from the electromagnetic self-interaction between 
these fermions, as those are expected to be small \cite{hossain2022methods}.
The generally invariant action for an electrically charged Dirac fermion $\psi$ 
coupled to an electromagnetic gauge field $\mathcal{A}_{\mu}$ is given by
\begin{equation}\label{DiracActionElectrons}
S = -\int\sqrt{-g}d^{4}x \bar{\psi} \Big[ie_{ \ a}^{\mu}
\gamma^{a} ( \mathcal{D}_{\mu} - ie \mathcal{A}_{\mu})
+ m \Big]\psi ~,
\end{equation} 
where $e$ denotes the electrical charge of the fermion.

\subsection{Partition function}

In order to evaluate the partition function, as earlier, we consider the 
external magnetic field to be along the $z$-direction and we choose the gauge 
field components to be $\mathcal{A}_{\mu} = (0,0,Bx,0)$. Therefore, within the 
box with the metric (\ref{MetricInBox}), we can reduce the Dirac action 
(\ref{DiracActionElectrons}) to the following form
\begin{equation}\label{ReducedActionP1}
S = -\int d^{4}x \bar{\psi}\Big[i\gamma^{0}\partial_{t} + 
e^{\Phi}(i\gamma^{k}\partial_{k} + m) + e^{\Phi}\gamma^{2} eB x\Big]\psi ~.
\end{equation}
As earlier, the partition function can be expressed as $\mathcal{Z} = 
\int\mathcal{D} \bar{\psi} \mathcal{D}\psi \ e^{-S^{\beta}}$ where $S^{\beta}$ 
denotes the Euclidean action corresponding to the action 
(\ref{ReducedActionP1}) and is given by
\begin{eqnarray}
S^{\beta} = \int_{0}^{\beta}d\tau\int d^{3}x \bar{\psi}\big[
&-& \gamma^{0}(\partial_{\tau} + \mu) + e^{\Phi} (i\gamma^{k}\partial_{k} + m)
\nonumber \\ \label{EuActionElectrons0}
&+& e^{\Phi} \gamma^2 e B x\big]\psi ~.
\end{eqnarray}
At thermal equilibrium, the Dirac field is subject to the \emph{anti-periodic} 
boundary condition $\psi(\tau,\x) = -\psi(\tau+\beta,\x)$ leading to the  
Matsubara frequencies $\omega_l = (2l+1) \pi/\beta$ where $l$ is an integer. 
Therefore, we can express the field $\psi$ in the Fourier domain as
\begin{equation}\label{FermionicFourierE}
\psi(\tau,\x) = \frac{1}{\sqrt{L_y L_z}} \sum_{l,k_y,k_z} 
~e^{-i (\omega_l\tau + k_y y + k_z z)} 
\psi_{l}(x,k_y,k_z)  ~,
\end{equation}
where $L_y$ and $L_z$ denote the length of the box in the $y$ and $z$ 
directions respectively. The equations (\ref{EuActionElectrons0}, 
\ref{FermionicFourierE}) then lead to
\begin{equation}\label{EuAction0II}
S^{\beta} = \sum_{l,k_y,k_z} \int dx ~\bar{\psi_{l}}~\beta
\left[ \tilde{\slashed{\mathcal{D}}} + \bar{m} \right] \psi_{l} ~,
\end{equation}
where $\tilde{\slashed{\mathcal{D}}} \equiv \gamma^a \tilde{\mathcal{D}}_a$ 
with 
\begin{equation}
\tilde{\slashed{\mathcal{D}}} = \gamma^{0}(i\omega_l - \mu) + \gamma^1 
(ie^{\Phi}\partial_{x}) + \gamma^2 e^{\Phi} (e B x + k_y)
+ \gamma^3 e^{\Phi} k_z ~.
\end{equation}
The partition function then can be expressed as
\begin{equation}\label{PartitionFunctionElec0}
\mathcal{Z} = \prod_{l,k_y,k_z} \text{det}[\beta(\tilde{\slashed{\mathcal{D}}} 
+ \bar{m})] ~.
\end{equation}
By using the property $\text{det}[\beta(\tilde{\slashed{\mathcal{D}}} + 
\bar{m})] = \text{det}[\gamma^5\beta(\tilde{\slashed{\mathcal{D}}} + 
\bar{m})\gamma^5] = \text{det}[\beta(-\tilde{\slashed{\mathcal{D}}} + 
\bar{m})]$ 
where $\gamma^5 \equiv i\gamma^{0}\gamma^{1} \gamma^{2}\gamma^{3}$, 
$(\gamma^5)^2 = \mathbb{I}$ and  $\{\gamma^{5},\gamma^{a}\} = 0$, one can show 
that
\begin{equation}\label{PartitionFunctionDeterminant}
\text{det}[\beta(\tilde{\slashed{\mathcal{D}}} + \bar{m})] = 
\text{det}[\beta^2(-\tilde{\slashed{\mathcal{D}}}^{2} + 
\bar{m}^2)]^{1/2} ~.
\end{equation}
Using the properties of the $\gamma^a$ matrices, we can express
\begin{equation}\label{DSlashed2m2}
-\tilde{\slashed{\mathcal{D}}}^{2} + \bar{m}^2 = (\omega_l + i\mu)^2
+ e^{2\Phi} \left[ H_{xy} - eB \Sigma_3 + k_z^2 + m^2\right] ~,
\end{equation}
where  $\Sigma_3 = \tfrac{i}{2}[\gamma^{1},\gamma^{2}] = \sigma_3 \otimes 
\mathbb{I}_2$ and 
\begin{equation}\label{ElectronHxy}
H_{xy} = -\partial_{x}^2 + (e B)^2 \left(x + 
\frac{k_y}{eB} \right)^2  ~.
\end{equation}
In order to evaluate the partition function (\ref{PartitionFunctionElec0}) we 
can compute the trace over the eigenstates of the operator $\Sigma_3$ with 
eigenvalues $2s$ where $s=\pm\tfrac{1}{2}$ and of the operator $H_{xy}$ with 
eigenvalues $(2n+1)|eB|$ where $n$ being non-negative integers. It 
leads to 
\begin{equation}\label{LogPartitionFunctionE0}
\ln\mathcal{Z} = \sum_{l,k_y,k_z,s,n} 
\ln \left[\beta^2 \{(\omega_l + i\mu)^2 + \varepsilon^2 \} \right] ~,
\end{equation}
where
\begin{equation}\label{ElectronEpsilon2}
\varepsilon^2 =  e^{2\Phi} \left[ eB (2n + 1 - 2s) + k_z^2 + m^2\right] ~.
\end{equation}
In the equation (\ref{ElectronEpsilon2}), for brevity of notation, the term
$|eB|$ is expressed as $eB$ and we shall use this notation henceforth.
One can carry out the summation over Matsubara frequencies $\omega_l$ (see 
\cite{hossain2021equation,hossain2022equation}) which leads to the following 
expression of the partition function
\begin{equation}\label{LogPartitionFunctionE2}
\ln\mathcal{Z} = \sum_{k_y,k_z,s,n} \left[ 
\ln\big(1 + e^{-\beta(\varepsilon - \mu)} \big) 
+ \ln\big(1 + e^{-\beta(\varepsilon + \mu)} \big)
\right] ~.
\end{equation}
In order to arrive at the equation (\ref{LogPartitionFunctionE2}), formally 
divergent terms such as the zero-point energy of fermions have been omitted. The 
first and the second terms in the equation (\ref{LogPartitionFunctionE2}) 
denotes the contributions from the \emph{particle} and the \emph{anti-particle} 
sectors respectively. Henceforth, we shall consider only the particle sector.

In the equation (\ref{ElectronEpsilon2}), we note that $\varepsilon$ is 
independent of $k_y$. However, in the equation (\ref{ElectronHxy}), $k_y$ 
shifts the origin of $x$-coordinate. Therefore, for a system of 
charged fermions in the given box, we must require $|k_y/eB| \le L_x/2$. By 
using the approximation $\sum_{k_y,k_z} = (L_yL_z)/(2\pi)^2 \int d k_y dk_z$, we 
can express the partition function for the particle sector as
\begin{equation}\label{LogPartitionFunctionE3}
\ln\mathcal{Z} =  \frac{eB V}{4\pi^2} \sum_{s,n} \int dk_z \ln\big(1 + 
e^{-\beta(\varepsilon - \mu)} \big) ~,
\end{equation}
where $V$ being the volume of the box. We note that in the partition 
function (\ref{LogPartitionFunctionE3}), we can replace the summation over 
the index $s$ and $n$ by a single summation over an index $\ell$ as follows
\begin{equation}\label{LogPartitionFunctionE4}
\ln\mathcal{Z} = \ln\mathcal{Z}_{0} + 2 \sum_{\ell=1} \ln\mathcal{Z}_{\ell} ~,
\end{equation}
where 
\begin{equation}\label{LogPartitionFunctionEll}
\ln\mathcal{Z}_{\ell} =  \frac{eB V}{2\pi^2} \int_{0}^{\infty} dk 
\ln\big(1 + e^{-\beta(\varepsilon_{\ell} - \mu)} \big) ~,
\end{equation}
with $\varepsilon_{\ell}^2 =  e^{2\Phi} \left[ 2(eB)\ell + k^2 + m^2 \right]$. 
The index $\ell$ here corresponds to the different Landau levels. From the 
equation (\ref{LogPartitionFunctionE4}), we note that the Landau levels, 
other than $\ell=0$, are \emph{doubly degenerate}.

By using the degeneracy condition of compact stars \emph{i.e.} $(\beta\mu) \gg 
1$, we can explicitly evaluate $\ln\mathcal{Z}_{\ell}$ as
\begin{equation}\label{LogPartitionFunctionEllFinal}
\ln\mathcal{Z}_{\ell} =  \frac{\beta V(eB)e^{-\Phi}}{4\pi^2}
\big[\mu\mu_{m\ell} 
- \bar{m}_{\ell}^2 \arcsinh (\mu_{m\ell}/\bar{m}_{\ell}) \big] ~,
\end{equation}
where $\bar{m}_{\ell}^2 = e^{2\Phi} \left[m^2 +  2(eB)\ell\right]$ and 
$\mu_{m\ell} = \sqrt{\mu^2 - \bar{m}_{\ell}^2}$. In order to ensure positive 
values for $\mu_{m\ell}$, we must restrict the summation over Landau levels up 
to an $\ell_{\max}$, given by
\begin{equation}\label{EllMax}
\ell_{max} =  \frac{\mu_{m}^2}{2(eB)e^{2\Phi}} ~ \text{with}~ 
\mu_{m} = \sqrt{\mu^2 - \bar{m}^2} ~.
\end{equation}
We can express the total partition function (\ref{LogPartitionFunctionE4}) as
\begin{equation}\label{LogPartitionFunctionESD}
\ln\mathcal{Z} = \ln\mathcal{Z}_{S}  + \ln\mathcal{Z}_{D} ~,
\end{equation}
where $\ln\mathcal{Z}_{S} \equiv \ln\mathcal{Z}_{\ell=0}$ represents the 
contributions from the \emph{singlet} Landau level and is given by
\begin{equation}\label{LogPartitionFunctionES}
\ln\mathcal{Z}_{S} =  \frac{(eB)\beta V e^{-\Phi}}{4\pi^{2}}
\left[\mu\mu_{m} - \bar{m}^{2}\arcsinh\left(\frac{\mu_{m}}{\bar{m}} 
\right)\right]  ~.
\end{equation}
On the other hand $\ln\mathcal{Z}_{D} \equiv 2 \sum_{\ell=1}^{\ell_{max}} 
\ln\mathcal{Z}_{\ell}$ represents the contributions from the \emph{doubly 
degenerate} Landau levels. With the aid of Poisson formula, by neglecting the 
oscillating part, we can evaluate it as $\ln\mathcal{Z}_{D} = 2 
\int_{1}^{\ell_{max}} d\ell \ln\mathcal{Z}_{\ell}$ \cite{tsintsadze2012relativistic} 
and it leads to
\begin{equation}\label{LogPartitionFunctionED}
\ln\mathcal{Z}_{D} =  \frac{\beta V e^{-3\Phi}}{24\pi^{2}}
\left[ 2\mu\mu_{m1}^{3} - 3\bar{m}_1^{2} \bar{\mu}_{m1}^2 \right] ~,
\end{equation}
where $\bar{m}_1 = \bar{m}_{\ell=1}$ and $\bar{\mu}_{m1}^2 = \mu\mu_{m1} 
-\bar{m}_1^{2} \arcsinh(\mu_{m1}/\bar{m}_1)$. It can be checked that in the 
absence of the magnetic field \emph{i.e.} as $B\to0$, $\bar{m}_1\to \bar{m}$, 
the total partition function (\ref{LogPartitionFunctionESD}) reduces exactly to 
the partition function of degenerate fermions as given in 
\cite{hossain2021equation}.

\subsection{Pressure and energy density}

Using the partition function (\ref{LogPartitionFunctionESD}), we can 
compute the number density of the fermions as
\begin{equation}\label{NumberDensityE}
n = \frac{1}{\beta V}\frac{\partial \ln\mathcal{Z}}{\partial\mu} 
= \frac{e^{-3\Phi}}{3\pi^{2}}\mu_{m1}^{3} 
+ \frac{eB e^{-\Phi}}{2\pi^{2}}\mu_{m} ~,
\end{equation}
where we have used the properties $(\partial \mu_m/\partial\mu) = \mu/\mu_m$,
$(\partial \mu_{m1}/\partial\mu) = \mu/\mu_{m1}$ and 
$(\partial \bar{\mu}_{m1}^2 /\partial\mu) = 2\mu_{m1}$.
For convenience, we now define $b = (3\pi^{2}/m^{3})$ and $B_{c} = (m^{2}/e)$ 
which then allows us to express $\mu_m$, up to $\mathcal{O}(B^2)$, as
\begin{equation}\label{MumUsingNumberDensity}
\mu_{m} = m e^{\Phi} \left[ (bn)^{1/3} + \frac{(B/B_c)}{2(bn)^{1/3}} \right] ~.
\end{equation}
We note that the constants $b$ and $B_c$ for charged fermions differ from 
the constants associated with neutrons. As earlier, we can compute the 
pressure as $P = (\beta V)^{-1} \ln\mathcal{Z}$ and express it as $P = (P_0 + 
P_B)$ where magnetic field independent part of the pressure is
\begin{eqnarray}\label{PressureP0}
P_0 =  \frac{m^4 e^{\Phi}}{24\pi^2} \left[ \sqrt{(b n)^{2/3} + 1}
\left\{2(b n) - 3(b n)^{1/3} \right\} \nonumber \right. \\
 + \left. ~3 \arcsinh\left\{ (b n)^{1/3} \right\} \right]  ~, 
\end{eqnarray}
and the magnetic field dependent part is
\begin{eqnarray}\label{PressurePB}
P_B =  \frac{m^4 e^{\Phi}}{12\pi^2} \frac{B}{B_c}
\left[ 3 \arcsinh\{ (b n)^{1/3} \} 
- \frac{(b n) + 3(b n)^{1/3}}{\sqrt{(b n)^{2/3} + 1}} \right] .
\end{eqnarray}
Similarly, the energy density $\rho$ can be expressed in terms of the 
partition function as $(\rho -\mu n)V = - 
(\partial\ln\mathcal{Z}/\partial\beta)$ leading to $\rho = \rho_0 + \rho_B$ 
where magnetic field dependent part of the energy density is
\begin{eqnarray}\label{EnergyDensity0E}
\rho_0 =  - P_0 + \frac{m^4 e^{\Phi}}{3\pi^2} (b n) \sqrt{(b n)^{2/3} + 1}  ~, 
\end{eqnarray}
and magnetic field dependent part is
\begin{eqnarray}\label{EnergyDensityBE}
\rho_B =  - P_B + \frac{m^4 e^{\Phi}}{6\pi^2} \frac{B}{B_c} 
\frac{(bn)}{\sqrt{(bn)^{2/3} + 1}} ~.
\end{eqnarray}
We again note that the EOS for an ensemble of electrically charged fermions 
under an external magnetic field and computed in the curved spacetime depends 
on the gravitational time dilation through the metric function $\Phi$, in 
addition to the magnetic field $B$. As expected, in the limit $B\to0$, the 
total pressure $P$ and energy density $\rho$ reduces to the standard 
expressions for degenerate fermions.

\begin{figure}
\includegraphics[height = 7cm, width = 9cm]{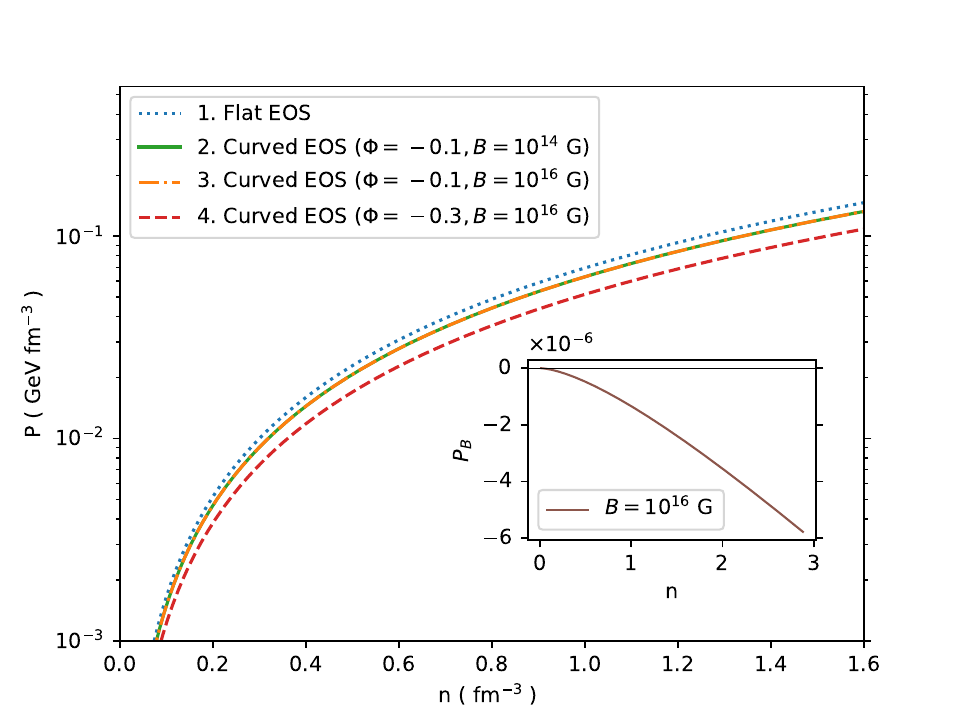}
\caption{Plot of the pressure $P$ exerted by an ensemble of degenerate protons 
under an external magnetic field $B$ as a function of number density $n$ for 
different kinematical values of metric function $\Phi$. The curves $2$ and $3$ 
with different values of magnetic field $B$ are almost indistinguishable. The 
change of pressure due to the non-zero magnetic field $B$ is shown in the inset 
plot using same units as the  main plot. It shows that effect of magnetic field 
on the pressure is rather small compared to the effect of gravitational 
time-dilation.}
\label{fig:pressure_comparison_P}
\end{figure}

\begin{figure}
\includegraphics[height = 7cm, width = 9cm]{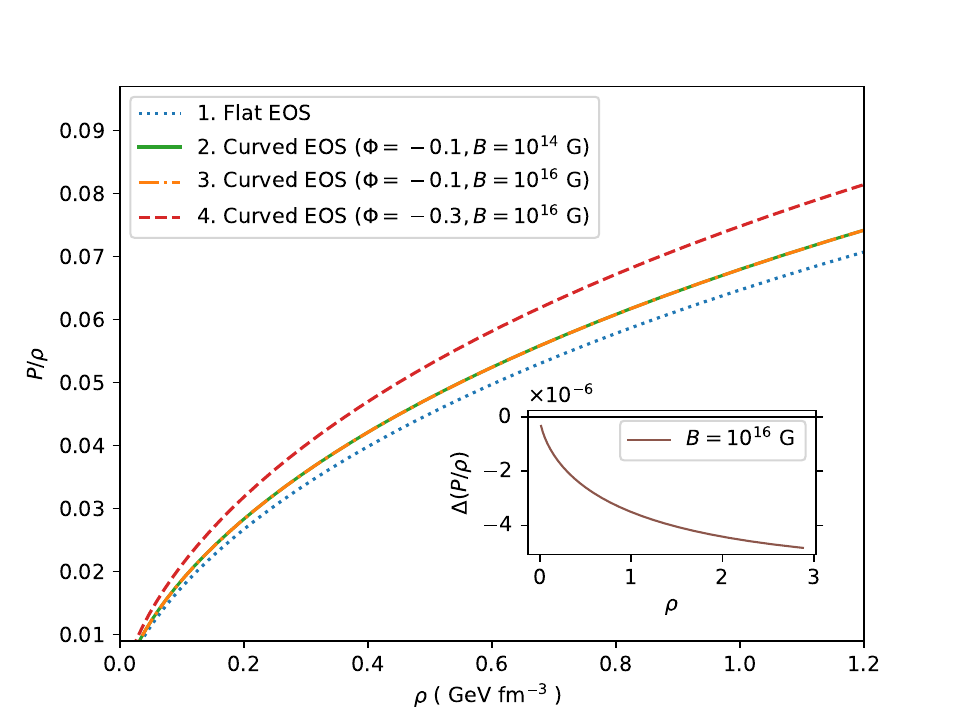}
\caption{Plot of the ratio $P/\rho$ as a function of the energy density $\rho$ 
for different values of magnetic field $B$ and metric function $\Phi$. As 
earlier, the curves $2$ and $3$ with different values of $B$ are nearly 
indistinguishable. The change of the ratio $P/\rho$ due to the magnetic field 
\emph{i.e.} $\Delta (P/\rho) \equiv (P/\rho)_{B} - (P/\rho)_{B=0}$ is shown in 
the inset plot. As earlier, we note that the effect of gravitational 
time-dilation on the equation of state is large compared to the effect of an 
external magnetic field. Nevertheless, due to formation of the Landau levels the 
effect of an external magnetic field on degenerate protons makes the EOS softer, 
unlike the case of neutrons.}
\label{fig:eos_ratio_P}
\end{figure}

The computed EOS in this section is valid for an ensemble of charged degenerate 
fermions in a compact star and in principle it could be used to describe 
degenerate protons and electrons in a neutron star as well as degenerate 
electrons in a white dwarf star. The different properties of the EOS for an 
ensemble of protons in a neutron star are plotted in the FIG. 
\ref{fig:pressure_comparison_P} and FIG. \ref{fig:eos_ratio_P}. In particular,  
the FIG. \ref{fig:eos_ratio_P} shows that unlike the case of neutrons, the 
effect of an external magnetic field on degenerate protons makes the 
corresponding EOS softer, essentially due to formation of the Landau levels 
which are bound states.

\subsection{Possible probe for de-confined quarks}

We note that for electrically charged fermions, an external magnetic 
field $B$ leads to $\mathcal{O}(B)$ corrections to the EOS. Further, 
these modifications are enhanced by the effects of curved spacetime and 
quantitatively these enhancements are dependent on the specific mass-radius 
curve of the star. Therefore, in principle one may use the presence of magnetic 
field as a possible probe for the existence of de-confined quarks 
which maybe present in the core of a neutron star (for example, see 
\cite{stangestars1986, ferrer2016exploring, rabhi2009quark, baym2018hadrons, 
franzon2016effects}). The quarks are known to be lighter compared to the 
nucleons. For example, the Up quark has mass, say $m_q$, of around $2.2$ MeV 
and it has electrical charge $e_q=2e/3$ which implies its characteristic magnetic 
field to be $B_c = m_q^2/e_q \sim 10^{15}$ Gauss. Therefore, if the core of a 
neutron star has de-confined quark degrees of freedom and it has magnetic field 
of around $10^{15}$ Gauss as indicated by observations then the EOS near the 
core of a neutron star should pick up a substantial corrections due to the 
magnetic field.

\section{Discussions}

In summary, in this article we have shown that for an ensemble of electrically 
neutral degenerate neutrons both magnetic field and gravitational time-dilation 
leads the EOS to become stiffer. However, for electrically charged fermions the 
magnetic field makes the EOS to become softer due to formation of the Landau 
levels. Nevertheless, the changes of EOS due to the gravitational time dilation 
is relatively stronger than the changes due to the observed strengths of 
magnetic field. We have shown that in presence of a non-zero magnetic field, 
a thin layer containing only spin-up neutrons would form at the boundary of a 
degenerate neutron star. Hence, a neutron star would acquire a non-zero 
magnetic moment which in turn would lead to an accretion of charged particles 
surrounding the star. Further, we have argued that a strong magnetic field 
can act like a possible probe for existence of de-confined quarks in the core 
of a neutron star where the effects of curved spacetime would enhance the 
modifications of the EOS.

\begin{acknowledgments}
SM is supported by SERB-Core Research Grant (Project RD/0122-SERB000-044). GMH 
acknowledges support from the grant no. MTR/2021/000209 of the SERB, Government 
of India.
\end{acknowledgments}

\bibliographystyle{apsrev}
\bibliography{eos_magnetic}

\end{document}